\documentstyle[graphics,prl,aps]{revtex}
\begin{document}
\twocolumn[\hsize\textwidth\columnwidth\hsize\csname@twocolumnfalse%
\endcsname
\title{Two Parameters for Three Dimensional Wetting Transitions}
\author{P.\ S.\ Swain\cite{email} and A.\ O.\ Parry}
\address{Department of Mathematics, Imperial College, 180 Queen's Gate, London SW7 2BZ, United Kingdom.}
\draft
\maketitle
\begin{abstract}
Critical effects at complete {\it and} critical wetting in three dimensions are studied using a coupled effective Hamiltonian $H \lbrack s({\bf y}), \ell({\bf y}) \rbrack$. The model is constructed via a novel variational principle which ensures that the choice of collective coordinate $s({\bf y})$ near the wall is optimal. We highlight the importance of a new wetting parameter $\Omega(T)$ which has a strong influence on critical properties and allows the status of long-standing Monte-Carlo simulation controversies to be re-examined.
\end{abstract}
\vskip2pc]

A suitable microscopic starting point for modelling fluid adsorption in systems with short ranged forces at a planar wall (situated at $z=0$) is the Landau-Ginzburg-Wilson (LGW) Hamiltonian \cite{wetting}
\begin{eqnarray}
H_{LGW}[m] &  = & \int d{\bf y} \left\{ \int_0^\infty \left[ \frac{K}{2} (\nabla m)^2 +\phi(m) \right]dz \right. \nonumber \\
& & \left.  + \frac{c m_1^2}{2}-m_1 h_1  \right\} \label{LGW}
\end{eqnarray}
based on a local magnetization order parameter $m({\bf y},z)$. Here $h_1$ and $c$ are the surface field and enhancement respectively. The bulk potential $\phi(m)$ yields coexistence between bulk phases $\alpha$ and $\beta$ (with corresponding equilibrium magnetizations $m_\alpha>0$ and $m_\beta<0$) for sub-critical temperatures $T<T_C$ and zero bulk field, $h=0$. The magnetization at the wall is denoted by $m_1({\bf y})=m({\bf y},z=0)$. The model is expected to show wetting behaviour so that for sufficiently large surface field $h_1(>0)$ and high (sub-critical) temperatures $T>T_W$ (where $T_W$ is the wetting temperature) the wall-$\beta$ interface is completely wet by the $\alpha$ phase when $h=0^-$. A simple surface phase diagram may then show two types of continuous wetting transition in which the mean thickness, $\langle \ell \rangle$, of adsorbed $\alpha$ phase diverges leading to large scale fluctuations in the position of the $\alpha \beta$ interface, characterised by the perpendicular and parallel correlation lengths, $\xi_\perp$ and $\xi_\parallel$. We distinguish between critical wetting transitions which occur for $T\rightarrow T_W^-$ with fixed surface field (or for fixed temperature and $h_1 \rightarrow h_1^{W-}$) and $h=0^-$, and complete wetting in which $\langle \ell \rangle$ diverges for $T \ge T_W$ (or $h_1 \ge h_1^W$) and $h \rightarrow 0^-$.

Beyond a mean field (MF) approximation, analysis based on (\ref{LGW}) has proved too difficult. Instead most authors use effective interfacial models based on a collective coordinate $\ell({\bf y})$ which only accounts for fluctuations in the position (height) of the $\alpha \beta$ interface. The energy cost of the configuration is given by the Hamiltonian
\begin{equation}
H[\ell({\bf y})] = \int d{\bf y} \left\{ \frac{\Sigma_{\alpha \beta}(\ell)}{2} [ \nabla \ell({\bf y})]^2 + W(\ell({\bf y})) \right\} \label{FJ}
\end{equation}
where we have included the position dependence of the stiffness coefficient $\Sigma_{\alpha \beta}(\ell)$ emphasized by Fisher and Jin (FJ) and not allowed for in the original capillary wave approach \cite{fisher2}. The direct interaction with the wall is modelled by the binding potential $W(\ell)$ which is sensitive to the details of interatomic forces \cite{binding}. Renormalization group (RG) predict that for dimension $d=3$ and short ranged forces the critical properties of both complete and critical wetting depend on the dimensionless parameter $\omega(T)=\frac{k_B T \kappa^2}{4\pi \Sigma_{\alpha \beta}}$ \cite{brezin}, with $\kappa$ the inverse (true) correlation length of the bulk $\alpha$ phase and $\Sigma_{\alpha \beta}\equiv\Sigma_{\alpha \beta}(\ell=\infty)$, the free interfacial stiffness. The wetting parameter takes a universal value $\omega_C \simeq 0.77$ near bulk $T_C$ \cite{fisher1}. As is well known, the predictions of strong non-universal criticality for critical wetting are not consistent with extensive Monte-Carlo work \cite{binder1} which only show MF behaviour, and imply $\omega(T) \simeq0$. On the other hand more recent simulations \cite{binder2} of complete wetting reveal an effective value of $\omega(T)$ much larger than expected.

Even within MF theory however the interfacial model (\ref{FJ}) is problematic. While it correctly predicts correlations in the magnetization near (and along) the $\alpha \beta$ interface (in agreement with calculations based on the LGW model) it does not describe those occurring through the wetting layer i.\ e.\ from the wall to the $\alpha \beta$ interface. To remedy this Boulter and Parry (BP) \cite{parry1,boulter} have introduced a {\it second} collective coordinate which models the relatively small magnetization fluctuations near the wall and results in a coupled effective Hamiltonian. For complete wetting, occurring for $T \gg T_W$, the nature of the second collective coordinate is unambiguous and may be efficiently modelled using an `interfacial-like' field $\ell_1({\bf y})$ which accounts for (small) local translations of the magnetization profile. Remarkably for $d=3$ the coupling of fluctuations appears to have a profound effect on the critical behaviour and increases the effective value of $\omega(T)$ in agreement with the latest Ising model simulations \cite{binder2}. This may be traced to the existence of a marginal operator in three dimensions corresponding to translations in the origin of the binding potential \cite{lipowsky}. However, the BP analysis is not quite fully quantitative (even for $T \gg T_W$) and becomes pathological for temperatures close to critical wetting ($T \simeq T_W$) where the magnetization profile becomes locally translationally invariant near the wall. The status of the second collective coordinate is then in doubt and the influence, if any, of the coupling of order parameter fluctuations near the wall and $\alpha \beta$ interface cannot be assessed satisfactorily.

In this paper we present details of a systematic and {\it fully quantitative} theory of three dimensional complete {\it and} critical wetting transitions, which reaffirms the important role played by coupling effects. There are two aspects of our work that we wish to emphasize. The first concerns the method by which the coupled effective Hamiltonian is derived from the underlying LGW model and relates to the choice of collective coordinate at the wall. Specifically we introduce a novel variational principle which ensures that the collective coordinate description most accurately approximates the functional integral over the phase space of magnetization configurations appearing in the partition function. We believe that this aspect of our work is of more general importance than the application to wetting described here. The second aspect of our analysis is the quantitative results for observables characterising critical and complete wetting transitions. These can now depend on a second dimensionless wetting parameter $\frac{k_B T \kappa^2}{4\pi \Sigma_{11}}$, involving the stiffness coefficient of the collective coordinate near the wall. Within our theory the value of this parameter at $T_W$ is explicitly determined as
\begin{equation}
\Omega(T) = \frac{\chi_{11}^{(\alpha)}}{4 \pi m_1^2 \xi_{w\alpha}^2} 
\end{equation}
with $\chi_ {11}^{(\alpha)}= k_B T \frac{\partial m_1}{\partial h_1}$ and $\xi_{w\alpha}$ are the surface susceptibility and (second moment) transverse correlation length of the surface layer at the wall-$\alpha$ interface, respectively. We estimate that for Ising-like systems $\Omega(T)$ is close to unity (above the roughening temperature) and also approaches a universal value $\Omega_C$ \cite{parry2} as $T\rightarrow T_C$. Using reliable scaling and RG results \cite{rowlinson} we predict that $\Omega_C \simeq \frac{8}{7} \omega_C$ yielding $\Omega_C \simeq 0.9$. The influence of $\Omega(T)$ allows a quantitative explanation of the hitherto anomalous simulations of Binder, Landau and co-workers \cite{binder1}. 

Our method of allowing order parameter fluctuations at the wall is a perturbative one in which we first consider magnetization configurations which do not cater for such effects. To begin therefore we follow FJ \cite{fisher2} and define the collective coordinate $\ell({\bf y})$ as the surface of fixed magnetization $m^X=0$. Next note that in order to construct the model Hamiltonian it is sufficient to consider the properties of planar constrained profiles \cite{fisher2} and we first focus on the profile $m_{FJ}(z)$ (written $m_\pi(z;\ell)$ in \cite{fisher2}) which minimizes $H_{LGW}[m]$ subject to the planar crossing criterion constraint 
\begin{equation}
m_{FJ}(\ell)=0	\label{con1} 
\end{equation}
The second step concerns the inclusion of order parameter fluctuations near the wall and for this we consider profiles which are similar to $m_{FJ}(z)$ but allow for local translations {\it and} enhancements of the magnetization. Note that these subtleties need not arise for the field $\ell({\bf y})$ for which translations of the profile always dominate. Specifically we consider an additionally constrained profile $m(z)$, defined as the minimum of (\ref{LGW}) subject to both the FJ condition (\ref{con1}) and 
\begin{equation}
m(z_1+s \sin \delta)=m_{FJ}(z_1)+\kappa m_\alpha s \cos \delta \label{con2}
\end{equation}
where $z_1$ is an arbitrary position close to the wall (chosen such that $0 \lesssim \kappa z_1 \ll 1$) which does not enter the final results.

\begin{figure}[h]
\begin{center}
\scalebox{0.5}{\includegraphics{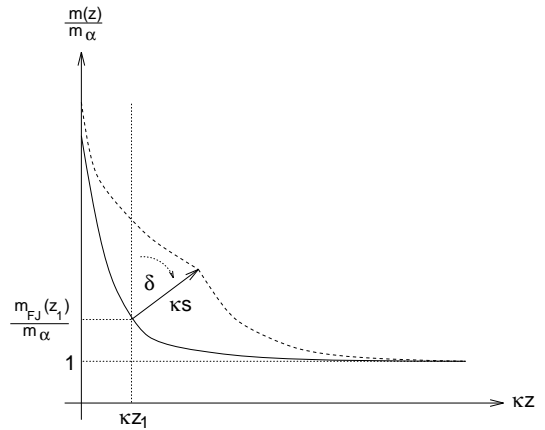}}
\end{center}
\caption{Detail of the planar magnetization profiles near the wall in scaled units. The broken curve shows $m(z)$  which incorporates a local enhancement and translation of the FJ profile (corresponding to the solid line). The proper coordinate $s$ and angle $\delta$ are shown.}
\end{figure}

The simple geometrical meaning of the proper collective coordinate $s$ and the variable $\delta$ is illustrated in Fig.\ 1. Note that for $\delta=\frac{\pi}{2}$ the fluctuation in the local magnetization represented by $s$ corresponds to a translation of the FJ profile. The coordinate $s({\bf y})$ is then the `interfacial-like' variable $\ell_1({\bf y})$ used by BP. However their description is clearly problematic if $m_{FJ}(z)$ is flat near the wall (as it is close to $T_W$) and hence translationally invariant. To overcome this problem we instead chose a value $\delta=\delta^*(\ell)$ which maximizes the linear response $ \left. \frac{\partial m}{\partial s} \right|_{s=0}$ of the FJ profile to a change in proper collective coordinate $s$. This may be written
\begin{equation}
\left. \frac{\partial m}{\partial s} \right|_{s=0} = \kappa m_\alpha \cos \delta - \frac{\partial m_{FJ}}{\partial z}(z_1) \sin \delta  \label{corr} 
\end{equation}
which implies that $\delta^*$ satisfies
\begin{equation}
\tan \delta^*(\ell)=- \frac{\frac{\partial m_{FJ}}{\partial z}(z_1)}{\kappa m_\alpha} 
\end{equation}
This variational choice of angle $\delta$ can be seen to maximize the `difference' between $m(z)$ and $m_{FJ}(z)$ for a given $s$, i.\ e.\ in Fig.\ 1, $s$ is `directed' along the normal to $m_{FJ}$ at $z=z_1$. The collective coordinate near the wall then provides the most efficient way of exploring the phase space of relevant magnetization configurations appearing in the partition function. One may compare with the `worst' choice of $\delta$ for which $s$ is `directed' along the tangent vector to $m_{FJ}(z)$ at $z_1$ --- variation of $s$ then leaves the FJ profile unchanged (to first order). The same criterion can also be derived algebraically by considering the structure of correlation functions \cite{parry2}. We suspect that similar ideas may be pertinent whenever a collective coordinate description is used to model low energy fluctuations. Considering our own problem this approach allows us to develop a coupled theory of both complete {\it and} critical wetting transitions which is not possible in the BP description. These phenomena are considered separately below.

(A) {\it Complete wetting}: The essential observation here is that in the approach to a complete wetting transition ($h\rightarrow0^-$ for $T>T_W$ or $h_1>h_1^W$) the angle $\delta^*$ is largely independent of $\ell$ and may be regarded as a function of $T$ and $h_1$. We find
\begin{eqnarray}
\delta^* & \simeq & \arctan \left[ \frac{h_1-c m_\alpha}{(c+\kappa) m_\alpha} \right] \nonumber \\
& \sim & \frac{h_1-h_1^W}{h_1^W} \mbox{\ \ \ \ \ for $h_1 \gtrsim h_1^W$}
\end{eqnarray}
where we have used $c m_\alpha(T) = h_1^W$. Deep in the complete wetting regime ($T \gg T_W$ or $h_1 \gg h_1^W$) $\delta^*$ approaches $\frac{\pi}{2}$ so that the proper collective coordinate is `interface-like' and the optimum model is basically the same as the earlier BP approach. We can therefore anticipate that the effective value of the wetting parameter is renormalised in accordance with this simpler theory \cite{parry1,boulter}. On the other hand as we decrease the surface field (or temperature) and consider complete wetting transitions closer to the critical wetting phase boundary the angle is small ($\delta^* \sim T-T_W$ or $\delta^* \sim h_1-h_1^W$) and the proper collective coordinate is `spin-like' (i.\ e.\ an enhancement of the profile in a vertical direction in Fig.\ 1). The absence of any interfacial or translation component implies that any increment to $\omega(T)$, associated with the coupling of interfacial fluctuations, vanishes. 

The effective Hamiltonian is constructed from the constrained profile (with  $\delta=\delta^*$) using well established techniques \cite{fisher2,parry1,parry2}. To calculate the critical exponents and amplitudes it is sufficient to ignore the position dependence of the stiffness matrix elements, in which case the coupled model is
\begin{eqnarray}
H[s,\ell] & = & \int d{\bf y} \left\{ \frac{\Sigma_{11}}{2} (\nabla s)^2 + \frac{\Sigma_{\alpha \beta}}{2} (\nabla \ell)^2 + \frac{r}{2} s^2 \right.  \nonumber \\
& & \biggl.  + W(\ell-s \sin \delta^*) \biggr\}
\end{eqnarray}
where $W(\ell)$ is basically the same function as that appearing in the CW model \cite{binding}. The parameter $r$ is related to the transverse correlation length of the wall-$\alpha$ interface by $\xi_{w \alpha}^2=\frac{\Sigma_{11}}{r}$ \cite{boulter,parry2}. The momentum cut-off of the $\ell({\bf y})$ field is the same as that in the CW model (i.\ e.\ $\Lambda_2 \lesssim \pi \kappa$ \cite{wetting}). Unlike the earlier coupled model which is problematic near the cross-over to critical wetting the cut-off $\Lambda_1$ for the lower field is well-behaved at all sub-critical temperatures. In particular near the critical wetting phase boundary $\Lambda_1$ is essentially identical to the cut-off of the LGW model ($\Lambda_0$ say) since the proper collective coordinate is `spin-like'. The influence of fluctuations on the critical singularities of the complete wetting transition is manifest in the critical amplitude \cite{parry1,boulter}
\begin{equation}
\theta= \stackrel{\rm lim}{\scriptstyle{h \rightarrow0}} \left\{ -\frac{\kappa \langle \ell \rangle}{\ln |h|} \right\}
\end{equation}
The RG analysis of the new Hamiltonian is very similar to that for the simpler theory \cite{boulter} and as before we find $\theta=1+\frac{\bar \omega}{2}$ but with $\bar{\omega}$, the renormalised value of the wetting parameter, given explicitly by
\begin{equation}
\bar{\omega}(T,h_1) = \omega(T) +\frac{\Omega(T_W)}{1+[\Lambda_0 \xi_{w\alpha}]^{-2}} \delta^{*2} + O(\delta^{*3}) \label{oe}
\end{equation}
where the higher order term is negative \cite{parry2}. Thus the optimized theory yields a precise perturbation expression for the surface field or temperature dependence of the renormalised wetting parameter. It follows that the increment to $\omega(T)$ vanishes quadratically as $T \rightarrow T_W^+$, a result which can not be derived from the BP model. This is illustrated in Fig.\ 2 and compares favourably with the available Ising model data (see Fig.\ 1 of \cite{parry3}). Note that for $T_W$ close to $T_C$ the second term in (\ref{oe}) reduces to $\Omega_C \left[ \frac{h_1-h_1^W}{h_1^W} \right]^2$ so that $\bar{\omega}$ has a {\it universal} expansion about $\omega_C$.

\begin{figure}[h]
\begin{center}
\scalebox{0.5}{\includegraphics{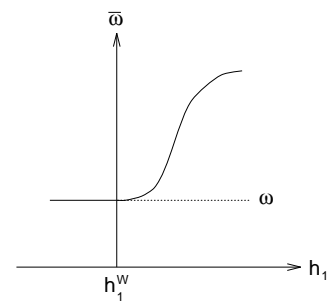}}
\end{center}
\caption{Schematic illustration of the effective value of the wetting parameter near a critical wetting transition (occurring at $h_1=h_1^W$). The coupling of fluctuations leads to an increment to the CW result $\omega$ which then vanishes quadratically as $h_1 \rightarrow h_1^{W+}$ i.\ e.\ for complete wetting transitions occurring close to the critical wetting phase boundary.}
\end{figure}

(B) {\it Critical wetting}: In many aspects the optimal coupled model yields results for critical wetting similar to the CW and FJ theories \cite{fisher2,brezin}. That is, if one ignores the position dependence of the stiffness coefficients, the correlation length critical exponent $\nu_\parallel$ is non-universal and dependent on $\omega$ \cite{brezin}. Including the position dependence of the stiffness coefficients there is a possibility that the transition is fluctuation induced first-order very similar to the stiffness instability mechanism of FJ. However there is behaviour associated with coupling effects which is not present in either of these models. This is most simply seen by calculating the Ginzburg criterion for the surface susceptibility $\chi_1=\frac{\partial m_1}{\partial h}$ and comparing the result with that of CW theory \cite{halpinhealy}. Along the critical isotherm ($T=T_W$ or $h_1=h_1^W$ and $h \rightarrow0^-$) $\chi_1$ should show cross-over from mean field-like behaviour ($\chi_1 \sim |h|^{-\frac{1}{2}}$) to the true asymptotic criticality ($\chi_1 \sim |h|^{-\frac{1}{2\nu_\parallel}}$ with $\nu_\parallel \simeq 3.7$ for $T_W \lesssim T_C$) when the correlation length $\xi_\parallel$ is close to a value satisfying
\begin{eqnarray}
& \omega(T_W) \left\{ \frac{1}{2} \ln [1+(\xi_\parallel \Lambda_2)^2]+\frac{1}{1+ (\xi_\parallel \Lambda_2)^2}-1 \right\}&  \nonumber \\
&  =1+\Omega(T_W) \ln[ 1+(\xi_{w\alpha} \Lambda_0)^2] &
\end{eqnarray}
Again this shows the presence of two wetting parameters in contrast to a CW approach \cite{halpinhealy} which is equivalent to $\Omega=0$. The upshot is that the true critical region is much smaller than previously expected. If we assume that both $\kappa^{-1}$ and $\xi_{w\alpha}$ are of order the (unit) lattice spacing of an Ising model (which is a reasonable estimate for the existing simulation studies \cite{binder1}) then the cross-over region corresponds to $\xi_\parallel \sim 60$ (lattice spacings) which is an {\it order of magnitude bigger} than the CW result. Following \cite{halpinhealy} we estimate that for the Ising model simulations the bulk field $h$ (written $\frac{H}{J}$ in \cite{binder1}) needs to be $\sim 6 \times 10^{-6}$ before non-MF behaviour is observed. This is much smaller than the range of fields studied and may well be beyond the limits of present day Monte-Carlo techniques. 

In summary, we have developed a precise quantitative theory of three dimensional wetting transitions, incorporating coupling effects. Deep in the complete wetting regime ($T \gg T_W$) our theory recovers the earlier model of BP but also allows a new discussion of cross-over effects from complete to critical wetting. In particular the increment to the effective value of the wetting parameter is predicted to decrease as $T \rightarrow T_W^+$ (or $h_1 \rightarrow h_1^{W+}$) in agreement with simulations \cite{binder2}. Applications to the critical wetting transition reveal a significant change to the Ginzburg criterion implying that MF behaviour for the susceptibility $\chi_1$ should remain valid until the transverse correlation length is extremely large. This appears to resolve long-standing problems associated with Ising model simulations \cite{binder1} which only find MF behaviour for this response function.

We are very grateful to Dr.\ C.\ J.\ Boulter for discussions and acknowledge support from the Engineering and Physical Sciences Research Council, United Kingdom.


\begin{references}
\bibitem{binding} The standard expression is $W(\ell)= \bar{h} \ell + a {\rm e}^{-\kappa \ell} + b {\rm e}^{-2\kappa \ell}$ where $\bar{h}$ is proportional to the bulk ordering field and  $b>0$. The parameter $a$ vanishes at the mean field critical wetting temperature and a hard wall constraint at the origin keeps $\ell>0$.
\end{references}
\end{document}